\documentclass[pre, aps, groupedaddress, showpacs, twocolumn]{revtex4}

\usepackage{amsmath,amsfonts,graphicx}

\newcommand{\Gf}{Green's function }
\newcommand{\Gfs}{Green's functions }

\begin{document}
\bibliographystyle{apsrev}

\title{Resonant Activation Phenomenon for \\  Non-Markovian Potential-Fluctuation Processes}

\author{Tom\'a\v s Novotn\'y}
\email[]{novotnyt@fzu.cz} \homepage[]{http://www.fzu.cz/~novotnyt}
\affiliation{Institute of Physics, Academy of Sciences of the
Czech Republic, Na Slovance 2, 182 21 Prague, Czech Republic}

\author{Petr Chvosta}
\email[]{chvosta@kmf.troja.mff.cuni.cz} \affiliation{Department of
Macromolecular Physics, Faculty of Mathematics and Physics,
Charles University, V~Hole\v{s}ovi\v{c}k\'ach 2, 180 00 Prague,
Czech Republic}

\date{August 11, 2000}

\begin{abstract}
We consider a generalization of the model by Doering and Gadoua to
non-Markovian potential-switching generated by arbitrary renewal
processes. For the Markovian switching process, we extend the
original results by Doering and Gadoua by giving a complete
description of the absorption process. For all non-Markovian
processes having the first moment of the waiting time
distributions, we get qualitatively the same results as in the
Markovian case. However, for distributions without the first
moment, the mean first passage time curves do not exhibit the
resonant activation minimum. We thus come to the conjecture that
the generic mechanism of the resonant activation fails for
fluctuating processes widely deviating from Markovian.
\end{abstract}


\maketitle

The resonant activation phenomenon first reported by Doering and
Gadoua \cite{doe-gad-prl-92} has attracted much attention. After
the famous seminal paper various models exhibiting this phenomenon
were considered and the conditions under which the phenomenon is
present were intensively studied \cite{reimann}. We consider a
simple generalization of the original model by Doering and Gadoua
to non-Markovian switching potentials generated by arbitrary
renewal processes. That is, we study the solutions of the
following stochastic differential equation
\begin{equation}
\label{model}
    \frac{d}{dt}{X}(t)= F Q(t) + \xi(t) \ ,
\end{equation}
where $\xi(t)$ is a Gaussian white noise process with
$\langle\xi(t)\xi(t')\rangle=2\delta(t-t')$ and $Q(t)$ is a
symmetric dichotomous noise jumping between $\pm 1$ according to a
renewal process generated by arbitrary waiting time density $f(t)$
with the distribution function $F(t)=\int_0^t d\tau\, f(\tau)$. We
study both the stationary as well as non-stationary renewal
processes. To make the process stationary we have to replace the
first waiting time density by $h(t)=\tfrac{1}{m_1}[1-F(t)]$ (the
distribution function $H(t)$), whit $m_1=\int_0^{\infty} dt\, t
f(t)$. In non-stationary cases we take $h(t)=f(t)$. The diffusion
described by the random process $X(t)$ takes place on a linear
segment $x\in(0,1)$ with the reflecting wall at $x=1$ and
absorbing wall at $x=0$.

First, we present the theory of the calculation of the
switching-averaged Green's functions. More precisely, the motion
within the safe domain $x\in(0,1)$ will be described by the
state-of-potential resolved conditional densities
$G_{\alpha\beta}(x,y;t)$, defined as
\begin{widetext}
\begin{equation}
    G_{\alpha\beta}(x,y;t)\,dx= \text{Prob}\{X(t)\in(x,x+dx)
    \text{ and }Q(t)=\alpha\,| \,X(0)=y\text{ and }Q(0)=\beta\}\ ,
\end{equation}
\end{widetext}
with $\alpha,\beta=\pm$. The subscripts specify the final
($\alpha$) and initial ($\beta$) state of the fluctuating
potential while the arguments $x,y$ give the final ($x$) and
initial ($y$) position of the diffusing particle. Because of the
absorbing boundary at $x=0$ the total probability in the safe
domain is not conserved. Instead, it gradually leaks out into the
boundary. It is helpful to consider auxiliary quantities called
boundary channels which describe the process of leaking of the
probability out of the safe domain into the absorbing boundary. We
may even distinguish two boundary channels according to the state
of the fluctuating potential at the moment of the absorption
event. In order to describe the dynamics of the channel filling
processes, we introduce the boundary channel occupation Green's
functions. These quantities are given by
\begin{equation}
    \pi_{\alpha\beta}(y;t)=\text{Prob}\{X(t)\in\text{B}_{\alpha}\,
    |\,X(0)=y\text{ and }Q(0)=\beta\}\ ,
\end{equation}
where the $\text{B}_{\pm}$ denote the corresponding boundary
channels. Again, the second subscript ($\beta$) relates to the
initial state of the potential while the first ($\alpha$)
specifies the boundary channel in question. It is convenient to
write these \Gfs in the form of 2-by-2 matrices denoted by the
boldface letters $\mathbf{G}(x,y;t)$, $\mathbf{\Pi}(y;t)$ in the
next.

Assume for a moment that the potential is static, i.e.\ it is
fixed in one of its two states. Then the above matrices which will
play a role of unperturbed quantities are diagonal with the form
$\mathbf{G}^{(0)}(x,y;t)=\text{diag}\bigl(G_{+}^{(0)}(x,y;t),G_{-}^{(0)}(x,y;t)\bigr)$,
$\mathbf{\Pi}^{(0)}(y;t)=\text{diag}\bigl(\pi_{+}^{(0)}(y;t),\pi_{-}^{(0)}(y;t)\bigr)$.
The \Gfs ${G}^{(0)}_{\pm}(x,y;t)$ are given simply by the
Fokker-Planck equations for the fixed potentials supplemented by
the boundary conditions. The boundary channel occupation
probabilities $\pi_{\pm}^{(0)}(y;t)$ are given by the global
probability conservation condition
\begin{equation}
\label{th:global-cons}
\pi_{\pm}^{(0)}(y;t)=1-\int_{0}^{1}dx\,G_{\pm}^{(0)}(x,y;t)\ ,
\end{equation}
which may be put with the help of the appropriate Fokker-Planck
equations in a local form reading
\begin{equation}
\label{th:local-cons}
\begin{split}
    \frac{\partial}{\partial t}\pi_{\pm}^{(0)}(y;t)&=
    \left.-\Bigl(-\frac{\partial}{\partial x}\pm
    F\Bigr)G_{\pm}^{(0)}(x,y;t)\right|_{x=0} \\
    &=\left.\frac{\partial}{\partial x}G_{\pm}^{(0)}(x,y;t)\right|_{x=0} \ .
\end{split}
\end{equation}
The last but one expression gives a clear physical insight as the
operator $-\frac{\partial}{\partial x}\pm F$ is the probability
current operator for the corresponding slope of the potential.
Namely, the rate of the channel filling process equals the
probability current into the absorbing boundary.

After these preparatory steps, let us focus on the construction of
the full \Gfs for fluctuating potentials. Basically, our procedure
consists of three steps. First, we shall assume an arbitrary fixed
sequence of the potential-switching events and we shall follow the
particle diffusion in the corresponding time dependent potential.
Secondly, we attribute to any such evolution the probability
weight of its realization. It is during this step that the
properties of the underlying potential-switching process enter the
calculation. Finally, we shall perform the averaging over the
complete set of mutually exclusive evolutions. The averaged
evolution is simply given by the sum over all possible evolutions
weighted by the corresponding probabilities. This is a brief
outline of the method of {\em construction of trajectories}
introduced by Chvosta and Reineker in \cite{chv-rei-pha-99}, where
also the full formalism can be found. The result of the procedure
in our case is
\begin{widetext}
\begin{equation}\label{th:konstrukce}
\begin{split}
  \mathbf{G}(t) &= (1-H(t))\mathbf{G}^{(0)}(t) + \int_0^t dt_1
    (1-F(t-t_1))\mathbf{G}^{(0)}(t-t_1)\cdot\mathbf{S}\cdot
    h(t_1)\mathbf{G}^{(0)}(t_1) \\
    &+ \int_0^t dt_1 \int_0^{t_1} dt_2 (1-F(t-t_1))\mathbf{G}^{(0)}(t-t_1)\cdot
    \mathbf{S}\cdot f(t_1-t_2)\mathbf{G}^{(0)}(t_1-t_2)\cdot\mathbf{S}\cdot h(t_2)\mathbf{G}^{(0)}(t_2)
    + \dotsb \ ,
\end{split}
\end{equation}
\end{widetext}
where the symbol `$\,\cdot\,$' denotes the matrix operator
multiplication, i.e.\ the matrix multiplication and the
integration over the internal spatial variables, and where the
matrix $\mathbf{S}=\bigl(\begin{smallmatrix} 0 & 1 \\ 1 & 0
\end{smallmatrix}\bigr)$ represents the switching event.
The full evolution is represented as a sum of processes with zero,
one, two, etc.\ potential switching events.

In the case of evaluation of the boundary channel occupations,
i.e.\ the boundary channel part of the \Gf $\mathbf{\Pi}(y;t)$,
the above procedure of the construction of trajectories yields the
proper whole set of mutually exclusive paths for the evolution of
$\frac{\partial}{\partial t}\mathbf{\Pi}(y;t)$, not for
$\mathbf{\Pi}(y;t)$ as one may see by a closer inspection.
Repeating the averaging procedure \eqref{th:konstrukce} for
$\frac{\partial}{\partial t}\mathbf{\Pi}(y;t)$ and bearing in mind
\eqref{th:local-cons} we derive this simple relation between the
boundary channel occupation and the safe domain parts of \Gfs
\begin{equation}
  \frac{\partial}{\partial t}\mathbf{\Pi}(y;t) = \left.
  \frac{\partial}{\partial x} \mathbf{G}(x,y;t)\right|_{x=0} \ .
\end{equation}
This formula expresses the local conservation law of probability
for the composite diffusion and potential-fluctuation process.

The convolution structure of \Gf \eqref{th:konstrukce} enables to
rewrite the above complicated time integrals structures via the
Laplace transform into a simple geometrical series which may be
formally summed up to the infinite order giving
\begin{subequations}\label{th:GF}
\begin{widetext}
\begin{equation}
\begin{split}
    \mathbf{G}(z) &= [(1-H)\mathbf{G}^{(0)}](z) + [(1-F)\mathbf{G}^{(0)}](z)\cdot
        \mathbf{S}\cdot[h\mathbf{G}^{(0)}](z) \\
        &+ [(1-F)\mathbf{G}^{(0)}](z)
        \cdot\mathbf{S}\cdot[f\mathbf{G}^{(0)}](z)\cdot\mathbf{S}\cdot[h\mathbf{G}^{(0)}](z)
        + \dots \\
        &= [(1-H)\mathbf{G}^{(0)}](z) + [(1-F)\mathbf{G}^{(0)}](z)\cdot
        (\mathbf{1} - \mathbf{S}\cdot[f\mathbf{G}^{(0)}](z))^{-1}
        \cdot\mathbf{S}\cdot[h\mathbf{G}^{(0)}](z) \ ,
\end{split}
\end{equation}
\end{widetext}
\begin{equation}
   \label{th:pi}
   \mathbf{\Pi}(y;z) = \frac{1}{z}\left.\frac{\partial}{\partial x}
        \mathbf{G}(x,y;z)\right|_{x=0} \ .
\end{equation}
\end{subequations}
Structures like $[f\mathbf{G}^{(0)}](z)=\int_0^{\infty} dt\,
e^{-zt} f(t)\mathbf{G}^{(0)}(t)$ mean a Laplace transform of the
product. Equations \eqref{th:GF} are our main result for \Gfs of
the composite process. Solving them we get the complete
information about the absorption process, i.e.\ the full
description of the time evolution of the probability captured in
the individual boundary channels.

In the following, we restrict ourselves mostly to a reduced
information concerning the boundary channel occupations. Namely,
we will consider the asymptotic boundary channel occupation
quantities defined as
\begin{equation}\label{th:ABCO}
    P_{\alpha\beta}(y) = \lim_{t \to\infty}\pi_{\alpha\beta}(y;t)
                       = \lim_{z \to 0} z\pi_{\alpha\beta}(y;z)
                            \ ,\quad\text{(ABCO)},
\end{equation}
and also the first moments of the boundary channels occupation
densities reading
\begin{equation}\label{th:tau}
  \tau_{\alpha\beta}(y) = \int_0^{\infty} dt \,
        t \frac{d\pi_{\alpha\beta}(y;t)}{dt} =
        \lim_{z\to 0}\frac{P_{\alpha\beta}(y)
        - z\,\pi_{\alpha\beta}(y;z)}{z}\ .
\end{equation}
These quantities are simply related to the mean first passage
times $\tau_{\pm}(y)$ for respective initial conditions $Q(0)=\pm$
by
\begin{equation}\label{th:MFPT}
  \tau_{\pm}(y) = \tau_{+\pm}(y) + \tau_{-\pm}(y)\ ,\quad\text{(MFPT)}.
\end{equation}

For the Markovian case, the waiting time distribution functions
are $f(t) = h(t) = \mu\, e^{-\mu t},\; 1-F(t) = 1-H(t) = e^{-\mu
t} = \frac{f(t)}{\mu} $. For any function $G(t)$ the following
identity holds $[fG](z) = \int_0^{\infty}dt e^{-zt} f(t)G(t) =
\mu\, G(z+\mu)$. We use these properties in \eqref{th:GF} to
obtain
\begin{equation}
\label{markov:GF}
  \mathbf{G}(z) = \begin{pmatrix}
                        \text{G}^0_+(z+\mu)^{-1} & -\mu \\
                        -\mu & \text{G}^0_-(z+\mu)^{-1}
                    \end{pmatrix}^{-1}
                  \ .
\end{equation}
Thus, for \Gf in the safe domain $\mathbf{G}(z)$, we simply get
the matrix Fokker-Planck equation valid for the Markovian
switching process (cf.\ Eq.\ (4) in \cite{doe-gad-prl-92}).

We may proceed further to analytically evaluate
$\pi_{\alpha\beta}(y;z)$ using \eqref{th:pi} from the knowledge of
the Markovian Green's function. With the help of identities being
satisfied by $\mathbf{G}(z)$ we come to the final analytic result
for the time evolution of the boundary occupation
$\mathbf{\Pi}(y;z)$ reading
\begin{equation}
\label{markov:reflexniPi}
   \mathbf{\Pi}(y;z) =
  \frac{1}{z}\,\mathbf{G}_R^{-1}(0,0;z)\cdot\mathbf{G}_R(0,y;z)
  \ ,\ \text{for $y\in(0,1)$ .}
\end{equation}
This expression uses the Markovian \Gf $\mathbf{G}_R(x,y;z)$ for
the diffusion in $(-\infty,1)$ with the reflecting wall at $x=1$
only.

\begin{figure}[t]
\includegraphics[width=80mm]{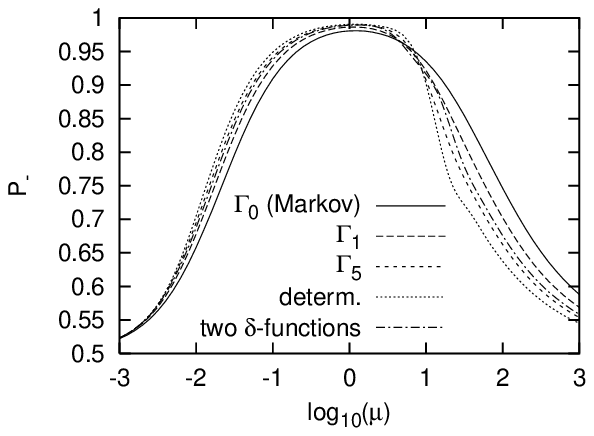}
\includegraphics[width=80mm]{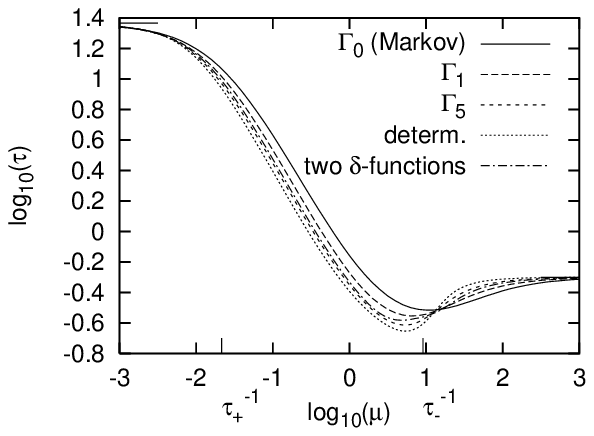}
\caption{ABCO and MFPT for waiting times distributions with fast
decay for $t\to\infty$, namely, $\Gamma_0$ (Markovian case) (solid
line), $\Gamma_1$ (long dashed line), $\Gamma_5$ (short dashed
line), deterministic switching (dotted line), and two
$\delta$-functions distribution (dot-dashed line). The
$\tau_{\pm}$ denote the mean first passage times for the
respective slopes of the potential. The limiting values of MFPT in
the static limit (the average of both $\tau$'s) and in the
infinitely fast switching limit (motion in the average potential)
are also shown by the bars.} \label{pic:moment}
\end{figure}

Instead of quoting the full rather involved result for
$\mathbf{\Pi}(y;z)$, we only present the physically transparent
expression for ABCO with the Doering-Gadoua initial condition
$y=1$
\begin{equation}
\label{markov:ABCOvzorce}
\begin{split}
   P_{\alpha\beta}(1) &= \Bigl[2\mu\cosh k + F^2 \Bigr]^{-1}
   \cdot\Bigl[\mu \cosh k + \alpha\beta\mu  \\
   & + \alpha\mu F\Bigl(1 - \frac{\sinh k}{k}\Bigr)  +
   \Bigl(\frac{\alpha+\beta}{2}\Bigr)^2 F^2 \Bigr] \ ,
\end{split}
\end{equation}
with $k=\sqrt{2\mu+F^2}$. One can easily see that the probability
conservation conditions for asymptotic times
$P_{+\beta}+P_{-\beta}=1$ are satisfied. We also verified that our
expression \eqref{markov:reflexniPi} leads to the famous result
for $\tau$ of Doering and Gadoua (10a-c). Moreover, we give here
the analytic expression for ABCO for the `minus' channel
$P_-(1)=\frac{P_{-+}(1)+P_{--}(1)}{2}$, a quantity analogous to
that depicted in Fig.\ 4 of \cite{doe-gad-prl-92} as a result of
the Monte Carlo simulation (for that calculation, DG used the
potential switching between $F_+=8 \text{ and } F_-=0$)
\begin{equation}\label{}
  P_-(1) = \frac{2\mu k\cosh k  + 2\mu F\sinh k  - 2\mu F k + k F^2}
        {2k\bigl(2\mu\cosh k+F^2\bigr)}\ .
\end{equation}
The curves of $P_-(1),\ \tau(1)$ as functions of $\mu$ for $F=8$
are plotted in Fig.\ \ref{pic:moment} for reference to be compared
with other results generated by non-Markovian switching.

Next, we present the numerical results for several non-Markovian
switching processes generated by various renewal processes
governed by waiting time probability densities $f(t)$ normalized
(except for one which cannot be normalized) so that the mean
switching time equals $\tfrac{1}{\mu}$, i.e.\
$\int_0^{\infty}dt\,t\,f(t)=\tfrac{1}{\mu}$. We evaluated the ABCO
of the `minus' channel $P_-(1) = \frac{P_{-+}(1)+P_{--}(1)}{2}$
and the MFPT
$\tau(1)=\frac{\tau_+(1)+\tau_-(1)}{2}=\frac{1}{2}\sum_{\alpha,\beta=\pm}
    \tau_{\alpha\beta}(1)$ for the symmetric initial condition
${\text{Prob}\{Q(0)=\pm\}=\frac{1}{2}}$ and $y=1$ considered also
in the Markovian case. To calculate these quantities we used a
numerical solution of \eqref{th:GF} employing the eigenmodes
expansion of the unperturbed Green's functions. It is of interest
to mention that for some specific waiting time distributions like
$\Gamma_n$ distributions used below, it would be possible to write
down and solve in an analytic form the equations for \Gfs
analogous to \eqref{markov:GF}.

\begin{figure}[t]
\includegraphics[width=80mm]{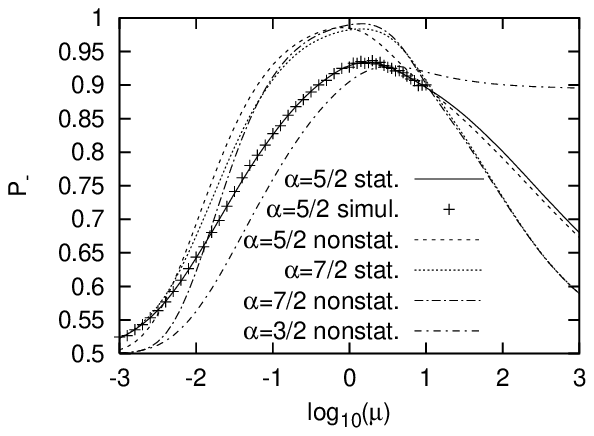}
\includegraphics[width=80mm]{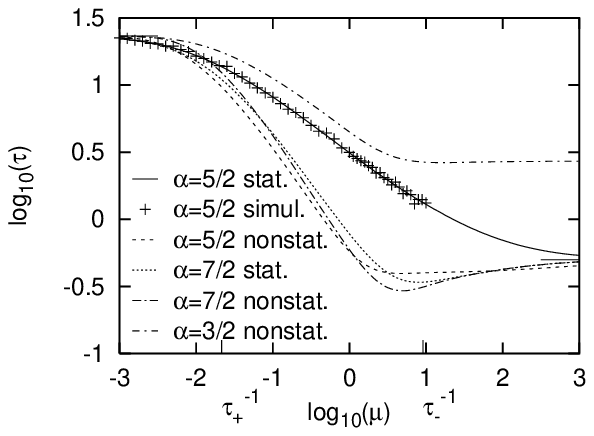}
\caption{ABCO for waiting time distributions without moments
(power laws $t^{-\alpha}$ in asymptotics). The respective curves
describes $\alpha=\frac{5}{2}$ stationary (solid line) and
non-stationary (dashed line) cases, $\alpha=\frac{7}{2}$
stationary (dotted line) and non-stationary (long dash-dotted
line) cases, and $\alpha=\frac{3}{2}$ non-stationary case (short
dash-dotted line). The results of Monte Carlo simulations for
$\alpha=\frac{5}{2}$ stationary case are also included.}
\label{pic:power}
\end{figure}

In the first set of pictures, Fig.\ \ref{pic:moment}, we plot the
results for the ABCO and MFPT for the stationary processes
generated by waiting time distributions decaying fast for
$t\to\infty$. The waiting time distributions used here cover the
$\Gamma_n$ distributions
$f_{\Gamma_n}(t)=\mu\tfrac{(n+1)^{(n+1)}}{n!}(\mu
t)^{n}\,e^{-(n+1)\mu t}$ with $n=0$ (Markov case), $1$ and $5$,
the delta-function distribution
$f_{\delta}(t)=\delta(t-\tfrac{1}{\mu})$ corresponding to the
deterministic switching process, and a two-delta-function
distribution
$f_{2\delta}(t)=\tfrac{1}{2}\,\delta(t-\tfrac{1}{2\mu})+\tfrac{1}{2}\,\delta(t-\tfrac{3}{2\mu})$.
One can see that the qualitative features of the Markovian case
are preserved even for the considered non-Markovian switching
potentials. Namely, both the wide resonant activation maximum in
the ABCO curve as well as the resonant activation minimum in the
MFPT curve are present in all cases with the shape changes
attributable to the various variances of the used probability
densities (compare $\Gamma_1$, two-delta-functions, and $\Gamma_5$
cases). We also performed the numerical simulations which fully
confirmed our results.

Further, we performed the calculations for the waiting time
distributions decaying slowly like power laws for large $t$, i.e.\
$f_{\alpha}(t) \propto t^{-\alpha}$ for large $t$ with
$\alpha=\tfrac{3}{2},\text{ }\tfrac{5}{2},\text{ }\tfrac{7}{2}$.
The exact expressions for these densities are: $f_{\alpha}(t)=
\tfrac{\mu}{\sqrt{2\pi}(\mu t)^{\alpha}} e^{-\tfrac{1}{2\mu t}}$
for $\alpha=\tfrac{3}{2}$,$\tfrac{5}{2}$ and $f_{7/2}(t)= \tfrac{3
\sqrt{3}}{\sqrt{2\pi\mu^5 t^7}}\, e^{-\tfrac{3}{2\mu t}}$. These
densities do not have higher order moments for $k \geq \alpha -1$.
The distributions chosen above do not have moments starting from
the first, the second, and the third, respectively. The results
for these distributions, together with their non-stationary
counterparts (for the $\alpha=\tfrac{3}{2}$ case, only the
non-stationary result exists), are shown in Fig.\ \ref{pic:power}.
Moreover, we also present there the results of the Monte Carlo
simulations for the $\alpha=\tfrac{5}{2}$ stationary case. One can
see that the results are qualitatively the same as in previous
cases for the both $\alpha=\tfrac{7}{2}$ cases (stat. and
non-stat.) and the $\alpha=\tfrac{5}{2}$ non-stationary case. On
the other hand, the $\alpha=\tfrac{5}{2}$ stationary case and the
$\alpha=\tfrac{3}{2}$ case are qualitatively different since the
resonant activation minimum in the MFPT curve is absent in these
two cases. This shows that whenever a waiting time density with
divergent first moment is involved (the stationary
$\alpha=\tfrac{5}{2}$ case has $h(t)$ with divergent first moment)
the generic resonant activation behaviour is spoiled. An analogous
behaviour in physically different context was found by Barkai and
Fleurov \cite{bar-fle-pre-97}.

To conclude we have presented the calculations of the resonant
activation phenomenon for non-Markovian switching potentials
generated by renewal processes with various waiting time
densities. We found that the results are qualitatively the same
for all switching processes except for those which are generated
by the waiting time densities with divergent first moment. For
those processes the resonant activation minimum in the MFPT curve
is not present. The method used for the calculations may be easily
extended for general potential profiles and different boundary
conditions.



\end{document}